\documentclass[preprint,showpacs,aps,superscriptaddress]{revtex4}

\usepackage{graphicx}
\usepackage{dcolumn}
\usepackage{bm}

\newcommand{\dd}{\mathrm{d}}
\newcommand{\eq}[1]{(\ref{#1})}

\newcommand{\bun}{\hat{\mathbf{b}}}

\newcommand{\zun}{\hat{\bm{\zeta}}}

\newcommand{\bB}{\mathbf{B}}
\newcommand{\bE}{\mathbf{E}}

\newcommand{\boldr}{\mathbf{r}}

\newcommand{\bk}{\mathbf{k}}

\newcommand{\bv}{\mathbf{v}}
\newcommand{\bV}{\mathbf{V}}

\newcommand{\sumtext}[1]{ \raise -1.3ex\hbox{${\mbox{\Large $\Sigma$}\atop{\scriptstyle #1}}$} }

\begin{document}


\title{Intrinsic rotation with gyrokinetic models}

\author{Felix I. Parra}
\affiliation{%
Plasma Science and Fusion Center, Massachusetts Institute of Technology, Cambridge, MA.%
}%
\author{Michael Barnes}
\affiliation{%
Plasma Science and Fusion Center, Massachusetts Institute of Technology, Cambridge, MA.%
}%
\author{Iv\'{a}n Calvo}
\affiliation{%
Laboratorio Nacional de Fusi\'{o}n, Asociaci\'{o}n EURATOM-CIEMAT, Madrid, Spain.%
}%
\author{Peter J. Catto}
\affiliation{%
Plasma Science and Fusion Center, Massachusetts Institute of Technology, Cambridge, MA.%
}%

\date{\today}

\begin{abstract}
The generation of intrinsic rotation by turbulence and neoclassical effects in tokamaks is considered. To obtain the complex dependences observed in experiments, it is necessary to have a model of the radial flux of momentum that redistributes the momentum within the tokamak in the absence of a preexisting velocity. When the lowest order gyrokinetic formulation is used, a symmetry of the model precludes this possibility, making small effects in the gyroradius over scale length expansion necessary. These effects that are usually small become important for momentum transport because the symmetry of the lowest order gyrokinetic formulation leads to the cancellation of the lowest order momentum flux. The accuracy to which the gyrokinetic equation needs to be obtained to retain all the physically relevant effects is discussed.
\end{abstract}

\pacs{52.30.Gz, 52.65.Tt}
\maketitle

\section{Introduction}

Tokamak plasmas can rotate toroidally to high speeds because they are axisymmetric. The presence of toroidal rotation is beneficial: it suppresses large-scale MHD instabilities \cite{devries96, bondeson97} and reduces the turbulent radial flux of particles, energy and momentum \cite{waltz07, mantica09, barnes11a, highcock10, parra11b, highcock11}. In a number of experiments (e.g. JET \cite{romanelli09} and DIII-D \cite{luxon02})  toroidal rotation is achieved by direct injection of momentum via neutral beams. In these cases the rotation can be close to sonic, resulting in turbulence suppression by velocity shear and a transition to regimes of reduced transport \cite{devries09a, devries09b, parra11b}. However, neutral beam injection of momentum is less effective in large, dense tokamak plasmas, and in particular, it is expected that the momentum input in ITER \cite{shimada07} will be small. This situation has driven a great deal of research on spontaneous, or intrinsic, rotation \cite{rice07}. Even in experiments with no momentum injection, it is often observed that the toroidal rotation is non-negligible. This intrinsic rotation shows non-trivial dependences on different plasma parameters. 

Intrinsic rotation can be in the direction of the current or against it, and often both signs are found in the same shots at the same time in different locations within the plasma \cite{degrassie07, parra11e}. The rotation at the edge is usually co-current and it is not correlated to the rotation in the core, i.e. it is possible to find pulses with almost identical rotation at the edge that will have opposite velocities in the core. Several experimental measurements indicate that the gradients in temperature and pressure are the drive for intrinsic rotation \cite{scarabosio06, solomon10, rice11a}. However, it is found that pulses with very similar density and temperature profiles have intrinsic rotation profiles that have the opposite sign, and it is possible to jump from co-current to counter-current rotation within the same pulse with slight changes in the density and the temperature. These rotation reversals seem to be related to a density threshold \cite{bortolon06, rice11b} in TCV \cite{tonetti91} and Alcator C-Mod \cite{hutchinson94}, but it is not clear what determines the direction of the rotation in other tokamaks such as JET \cite{nave11}. RF heating and current drive complicate the picture even more because Ion Cyclotron Resonance Heating (ICRH) \cite{rice99, eriksson09, lin09} tends to drive co-current rotation, and Electron Cyclotron Resonance Heating (ECRH) drives counter-current rotation \cite{degrassie07, mcdermott11}. Lower Hybrid (LH) current drive can drive co- or counter-current rotation depending on the situation \cite{incecushman09, eriksson09, podpaly11}. Any modeling effort needs to consider several effects and must allow them to compete in order to obtain the variety of intrinsic rotation profiles observed in experiments.

In this article we discuss the features required for a self-consistent model of intrinsic rotation. Since rotation is determined by the radial flux of toroidal angular momentum, any model of intrinsic rotation will necessarily have to include the main cause of radial transport: the turbulence. The need to model the turbulence makes the problem particularly challenging because turbulence is a rich and complex phenomenon, and in tokamak plasmas is kinetic in nature. The problem is simplified by employing gyrokinetics \cite{catto78}, a reduced kinetic model adequate for low frequency fluctuations. Gyrokinetics averages over the gyrofrequency, which is too fast to matter for the fluctuations, and at the same time keeps wavelengths that are comparable to the gyroradius. To do this, an expansion in the small parameter $\rho_\ast = \rho_i/L \ll 1$ is performed, and terms of higher order neglected. Here $\rho_i$ is the ion gyroradius and $L$ is the typical size of the machine of interest. 

The problem with modeling intrinsic rotation is that the momentum flux that determines the intrinsic rotation profiles can only be obtained by solving for small corrections to the lowest order gyrokinetic treatment. This is necessary because of a symmetry present in the lowest order gyrokinetic system, first proposed in the context of quasilinear estimates for turbulent transport \cite{peeters05}. The complete symmetry of the non-linear set of gyrokinetic equations has only been recently derived and confirmed with gyrokinetic simulations \cite{sugama11, parra11c}. Because this symmetry was only recently identified, the turbulence community had not worried about the order to which the expansion in $\rho_\ast$ had to be carried out to obtain the correct radial flux of toroidal angular momentum. The order in $\rho_\ast$ to which is necessary to solve the gyrokinetic equation was obtained in \cite{parra10a}, where it was assumed that  there is a cancellation in the radial flux of momentum due to up-down symmetry.

One of the main conclusions of \cite{parra10a} is that none of the existing simulations of turbulent momentum transport contain all the physical phenomena that can generate intrinsic rotation. How far the codes are from a comprehensive and complete model depends on the approach. Some codes use fluid equations to evolve the profiles of density, temperature and rotation \cite{candy09, barnes10b}, and the gyrokinetic equation is solved in a $\delta f$ code \cite{dorland00, candy03a, chen03, dannert05, peeters09c} only to obtain the self-consistent fluctuations to then calculate the radial fluxes of particles, energy and toroidal angular momentum. In these simulations both the fluid equation for toroidal angular momentum and the gyrokinetic equation solved in the $\delta f$ code have to be modified to include effects that are an order higher in $\rho_\ast$. The problem is different for simulations that solve for the full distribution function and do not use fluid equations to evolve the density, the temperature and the rotation \cite{grandgirard06a, xu07, chang08, heikkinen08}. In these codes the rotation profile is obtained from a moment of the distribution function, and the transport equations that were used in the other approach are assumed to be contained correctly by the gyrokinetic Fokker-Planck equation for the full distribution function. It is far from obvious that the gyrokinetic equation describes correctly the evolution of density, temperature and velocity because the equation for the full distribution function is a gyrokinetic equation and it is truncated to some low order in $\rho_\ast$. This limitation was pointed out in a series of articles \cite{parra08, parra09b, parra10b, parra10c}, and it was concluded that it is necessary to keep many more terms in the gyrokinetic equation than the terms retained in existing simulations to recover the correct conservation equation for the toroidal angular momentum.

In this article we will review all these results and explain how they apply to the modeling of intrinsic rotation. In Section~\ref{section:intrinsic} we will discuss the approach that uses fluid equations because it is the most intuitive path to obtain the order to which the equations have to be expanded. In Section~\ref{section:fullf} we discuss the requirements on the approaches that do not use fluid equations, and in particular, we derive the implications for collisionless models that employ a phase-space Lagrangian gyrokinetic formalism. Finally, in Section~\ref{section:discussion} we close with a discussion that includes possible future approaches to the problem.

\section{Intrinsic rotation} \label{section:intrinsic}

In most tokamaks the time between ion-ion collisions is shorter than the transport time scale. As a result \cite{parra09b, parra10a}, the ion velocity is given by the neoclassical formula \cite{hinton76}
\begin{equation} \label{eq:velocity}
\bV_i = \Omega_\zeta (\psi) R \zun + \frac{\mu_i c I}{Z e \langle B^2 \rangle_\psi} \frac{\partial T_i}{\partial \psi} \bB,
\end{equation}
where the toroidal rotation angular velocity is
\begin{equation} \label{eq:rotation}
\Omega_\zeta (\psi) = - \frac{c}{Zen_i} \left ( \frac{\partial p_i}{\partial \psi} + Zen_i \frac{\partial \phi_0}{\partial \psi} \right ),
\end{equation}
and $\mu_i$ is a numerical factor that depends on the geometry of the flux surface and the collisionality $R \nu_{ii}/v_{ti}$. Here $n_i$, $\bV_i$, $T_i$ and $p_i = n_i T_i$ are the ion density, average velocity, temperature and pressure, $\phi_0$ is the lowest order electrostatic potential, $v_{ti} = \sqrt{2T_i/m_i}$ is the ion thermal speed, $\nu_{ii}$ is the ion-ion collision frequency, $Ze$ and $m_i$ are the ion charge and mass, $R$ is the major radius, $\zun$ is the unit vector in the toroidal direction, $\psi$ is the poloidal magnetic flux and has the role of radial coordinate, $\zeta$ is the toroidal angle, $\bB = I \nabla \zeta + \nabla \zeta \times \nabla \psi$ and $B$ are the magnetic field and its magnitude, $I = R B_\zeta$, $B_\zeta$ is the toroidal component of the magnetic field, $\nabla \zeta = \zun/R$, and $e$ and $c$ are the proton charge and the speed of light. To obtain the rotation $\Omega_\zeta$ we need to solve the conservation of toroidal angular momentum, given by
\begin{equation} \label{eq:momentum transport}
\frac{\partial}{\partial t} \left ( V^\prime \left \langle R n_i m_i \bV_i \cdot \zun \right \rangle_\psi \right ) = - \frac{\partial}{\partial \psi}  \left ( V^\prime \Pi \right ) + S_\zeta.
\end{equation}
Here $\langle \ldots \rangle_\psi = (V^\prime)^{-1} \int \dd\theta \, \dd\zeta\, (\bB \cdot \nabla \theta)^{-1} (\ldots)$ is the flux surface average, $V^\prime \equiv \dd V/\dd \psi = \int \dd\theta \, \dd\zeta\, (\bB \cdot \nabla \theta)^{-1}$ is the volume contained between two contiguous flux surfaces, $\theta$ is the poloidal angle,  $S_\zeta$ is a momentum source and
\begin{eqnarray}
\Pi = \Bigg \langle \Big \langle R m_i \int \dd^3v\, f_i (\bv \cdot \zun) (\bv \cdot \nabla \psi) \nonumber\\ - \frac{1}{4\pi} R (\bB \cdot \zun) (\bB \cdot \nabla \psi) \Big \rangle_\psi \Bigg \rangle_\mathrm{T}
\end{eqnarray}
is the radial flux of toroidal angular momentum, where $f_i (\boldr, \bv, t)$ is the ion distribution function, and $\langle \ldots \rangle_\mathrm{T} = (\Delta t)^{-1} (\Delta \psi)^{-1} \int_{\Delta t} \dd t \int_{\Delta \psi} \dd\psi (\ldots)$ is the coarse grain average over a time interval $\Delta t$ and a radial segment $\Delta \psi$ to average over the turbulent fluctuations. The radial flux of toroidal angular momentum has two main components: the motion of particles that transfers toroidal angular momentum from one flux surface to the next, given by $R m_i \int \dd^3v\, f_i (\bv \cdot \zun) (\bv \cdot \nabla \psi)$, and the Maxwell stress due to the fluctuating magnetic field $(4\pi)^{-1} R (\bB \cdot \zun) (\bB \cdot \nabla \psi)$ (note that the non-fluctuating part of the magnetic field satisfies $\langle \bB \rangle_\mathrm{T} \cdot \nabla \psi = 0$). In general, for low $\beta$ plasmas the Maxwell stress contribution is small compared to the radial flux of momentum due to the particles. For the rest of this article we will neglect it. We will also assume that the electric field is electrostatic $\bE = -\nabla \phi$. These assumptions are easily relaxed and do not change the final result.

Even in the absence of momentum injection, it is observed that tokamaks have a sizable rotation. To study this intrinsic rotation we need to find a non-zero solution to \eq{eq:momentum transport} for $S_\zeta = 0$. If in addition we consider a steady state, we are left with the equation $\partial ( V^\prime \Pi)/\partial \psi = 0$ that implies that $V^\prime \Pi$ is a constant. By imposing regularity at the magnetic axis, where $\nabla \psi = 0$, the final equation for the intrinsic rotation is that there is no net momentum flux through any flux surface, i.e.
\begin{equation}
\Pi [ \Omega_\zeta (\psi); n_e (\psi), T_i (\psi), T_e (\psi), ...] = 0.
\end{equation}
Thus, to solve for the profile of intrinsic rotation we need to obtain the functional dependence of $\Pi$ on $\Omega_\zeta (\psi)$ and then solve the equation $\Pi = 0$.

The radial flux of momentum $\Pi$ is dominated by turbulence. In the following sections we will treat the turbulence by expanding in the small parameter $\rho_\ast = \rho_i/L \ll 1$. In subsection~\ref{subsection:ordering} we review the characteristics of the turbulent fluctuations in a tokamak and we use it to split the ion distribution function and the electrostatic potential into different pieces. In subsection~\ref{subsection:momentum flux} we write the radial flux of toroidal angular momentum in a specific form that in combination with the ordering of subsection~\ref{subsection:ordering} will give a complete picture of the necessary pieces for momentum transport. We will find that one of the pieces in the radial flux of momentum seems to dominate. However, in subsection~\ref{subsection:symmetry} we are able to show using a symmetry of the system that the apparently large term in the radial flux of momentum vanishes and we need to keep the other terms. We discuss the implications for intrinsic rotation in subsection~\ref{subsection:consequences}.

\subsection{Ordering} \label{subsection:ordering}

In tokamaks the turbulent fluctuations are a factor of $\rho_\ast \ll 1$ smaller than the background quantities. Their characteristic frequency is of the order of $v_{ti}/L$, and their characteristic wavelength is of the order of the ion gyroradius $\rho_i$. The scale length of the background quantities is much longer, of the order of the size of the device $L$, and they evolve on a much longer transport time scale $\tau_E \sim \rho_\ast^{-2} L/v_{ti} \gg L/v_{ti}$. Thus, it is useful to think about the distribution function as composed of a long wavelength piece $F_i$ and a short wavelength turbulent component $f_i^\mathrm{tb}$, i.e.
\begin{equation}
f_i = F_i (\psi, \theta, \bv, t) + f_i^\mathrm{tb} (\boldr, \bv, t).
\end{equation}
Similarly, the electrostatic potential can be split into a long wavelength piece $\Phi$ and a fluctuating piece $\phi^\mathrm{tb}$, leading to
\begin{equation}
\phi = \Phi (\psi, \theta, t) + \phi^\mathrm{tb} (\boldr, t).
\end{equation}

The long wavelength piece of the distribution function that contains the background density, temperature and velocity is axisymmetric ($\partial/\partial \zeta = 0$) and is composed of pieces of different order in $\rho_\ast$, giving
\begin{equation}
F_i = f_{Mi} + F_{i1}^\mathrm{nc} + F_{i2} + \ldots,
\end{equation}
where $f_{Mi}$ is the lowest order Maxwellian, $F_{i1}^\mathrm{nc} \sim \rho_\ast f_{Mi}$ is the next order correction that contains the neoclassical corrections \cite{hinton76}, and $F_{i2} \sim \rho_\ast^2 f_{Mi}$ is the second order correction that has to do with both neoclassical processes and the turbulence. In \cite{parra10a} these different pieces are calculated employing a simplifying expansion in the ratio of the poloidal component of the magnetic field $B_p$ and the total magnetic field $B$, $B_p/B \ll 1$. The equations for the complete $F_{i2}$ without the expansion in $B_p/B \ll 1$ are given for the first time in \cite{parra11g}. 

The turbulent pieces of the potential and the distribution function have perpendicular wavelengths comparable to the ion gyroradius, $k_\bot \rho_i \sim 1$, but the parallel gradients are much smaller, $k_{||} L \sim 1$. This anisotropy is due to the critical balance between the characteristic frequency of the parallel motion $k_{||} v_{ti}$ and the characteristic frequency of the perpendicular drifts $k_\bot \bv_E^\mathrm{tb} \sim k_\bot \rho_\ast v_{ti}$, where $\bv_E^\mathrm{tb} = - (c/B) \nabla \phi^\mathrm{tb} \times \bun \sim \rho_\ast v_{ti}$ is the turbulent $\bE \times \bB$ drift \cite{barnes11b}. The fact that the characteristic perpendicular size of the eddies is much smaller than the characteristic size of the tokamak suggests the use of a local approximation in which the fluctuations are treated as an eikonal, i.e.
\begin{equation}
\phi^\mathrm{tb} (\boldr, t) = \sum_{\bk_\bot} \underline{\phi}^\mathrm{tb} (\psi, \theta, k_\psi, k_\alpha, t) \exp ( i \bk_\bot \cdot \boldr )
\end{equation}
and
\begin{equation}
f_i^\mathrm{tb} (\boldr, \bv, t) = \sum_{\bk_\bot} \underline{f_i}^\mathrm{tb} (\psi, \theta, k_\psi, k_\alpha, \bv, t) \exp ( i \bk_\bot \cdot \boldr ),
\end{equation}
where the perpendicular wavenumber is
\begin{equation}
\bk_\bot = k_\psi \nabla \psi + k_\alpha \nabla \alpha \sim \frac{1}{\rho_i}.
\end{equation}
Here $i = \sqrt{-1}$, and $\alpha$ is a magnetic coordinate perpendicular to the magnetic field and parallel to the flux surface defined by $\bB = \nabla \alpha \times \nabla \psi$. The fluctuations have a fast spatial variation associated with $k_\bot$ and a much slower poloidal and radial variation. The slow poloidal variation is obtained for $\psi$ and $\alpha$ fixed, and has to do with the long correlation lengths along magnetic field lines. The fluctuations $\phi^\mathrm{tb}$ and $f_i^\mathrm{tb}$ can be expanded in $\rho_\ast$ as well, giving
\begin{equation}
\phi^\mathrm{tb} = \phi^\mathrm{tb}_1 + \phi^\mathrm{tb}_2 + \ldots
\end{equation}
and
\begin{equation}
f_i^\mathrm{tb} = f_{i1}^\mathrm{tb} + f_{i2}^\mathrm{tb}+ \ldots,
\end{equation}
where $e \phi_1^\mathrm{tb}/T_e \sim f_{i1}^\mathrm{tb}/f_{Mi} \sim \rho_\ast$ and $e \phi_2^\mathrm{tb}/T_e \sim f_{i2}^\mathrm{tb}/f_{Mi} \sim \rho_\ast^2$.

\subsection{Radial flux of toroidal angular momentum} \label{subsection:momentum flux}

We will see soon that the radial flux of toroidal angular momentum $\Pi$ has to be calculated up to order $\rho_\ast^3 p_i R |\nabla \psi|$ to describe intrinsic rotation. In \cite{parra10a} a convenient form of the radial flux of toroidal angular momentum is calculated using moments of the Fokker-Planck equation. The convenience of this expression is discussed in Appendix B of \cite{parra11d}. To summarize the result found in \cite{parra10a}, the radial flux of momentum is made of two pieces
\begin{equation}
\Pi = \Pi_{-1} + \Pi_0.
\end{equation}
The piece $\Pi_{-1}$ is formally of order $\rho_\ast^2 p_i R |\nabla \psi|$ and is given by
\begin{eqnarray} \label{eq:pi-1}
\Pi_{-1} = \left \langle \left \langle m_i c R \frac{\partial \phi_1^\mathrm{tb}}{\partial \zeta} \int \dd^3v\, f_{i1}^\mathrm{tb} (\bv \cdot \zun) \right \rangle_\psi \right \rangle_\mathrm{T} \nonumber\\ - \left \langle \frac{m_i^2 c R^2}{2 Z e} \int \dd^3 v\, C_{ii}^{(\ell)} [ F_{i1}^\mathrm{nc} ] (\bv \cdot \zun)^2 \right \rangle_\psi,
\end{eqnarray}
where $C_{ii}^{(\ell)} [ f ]$ is the linearized ion-ion collision operator. The contribution $\Pi_0$ is of order $\rho_\ast^3 p_i R |\nabla \psi|$ and is given by
\begin{eqnarray} \label{eq:pi0}
\Pi_0 = \left \langle \left \langle m_i c R \frac{\partial \phi_2^\mathrm{tb}}{\partial \zeta} \int \dd^3v\, f_{i1}^\mathrm{tb} (\bv \cdot \zun) \right \rangle_\psi \right \rangle_\mathrm{T} \nonumber\\ + \left \langle \left \langle m_i c R \frac{\partial \phi_1^\mathrm{tb}}{\partial \zeta} \int \dd^3v\, f_{i2}^\mathrm{tb} (\bv \cdot \zun) \right \rangle_\psi \right \rangle_\mathrm{T} \nonumber\\ - \left \langle \frac{m_i^2 c R^2}{2 Z e} \int \dd^3 v\, C_{ii}^{(\ell)} [ F_{i2} ] (\bv \cdot \zun)^2 \right \rangle_\psi \nonumber\\ - \left \langle \frac{m_i^2 c R^2}{2 Z e} \int \dd^3 v\, C_{ii}^{(n\ell)} [ F_{i1}^\mathrm{nc}, F_{i1}^\mathrm{nc} ] (\bv \cdot \zun)^2 \right \rangle_\psi \nonumber\\ - \left \langle \left \langle \frac{m_i^2 c R^2}{2 Z e} \int \dd^3 v\, C_{ii}^{(n\ell)} [ f_{i1}^\mathrm{tb}, f_{i1}^\mathrm{tb} ] (\bv \cdot \zun)^2 \right \rangle_\psi \right \rangle_\mathrm{T} \nonumber\\ + \frac{1}{V^\prime} \frac{\partial}{\partial \psi}  V^\prime \left \langle \left \langle \frac{m_i^2 c^2 R^2}{2 Z e} \frac{\partial \phi_1^\mathrm{tb}}{\partial \zeta} \int \dd^3v\, f_{i1}^\mathrm{tb} (\bv \cdot \zun)^2 \right \rangle_\psi \right \rangle_\mathrm{T} \nonumber\\ - \frac{1}{V^\prime} \frac{\partial}{\partial \psi} V^\prime \left \langle \frac{m_i^3 c^2 R^3}{6 Z^2 e^2} \int \dd^3 v\, C_{ii}^{(\ell)} [ F_{i2} ] (\bv \cdot \zun)^3 \right \rangle_\psi,
\end{eqnarray}
where $C_{ii}^{(n\ell)} [ f, g ]$ is the complete bilinear ion-ion collision operator. If $\Pi_{-1}$ does not vanish, it dominates and it is unnecessary to compute the piece $\Pi_0$ and hence the second order corrections $F_{i2}$, $f_{i2}^\mathrm{tb}$ and $\phi_2^\mathrm{tb}$. However, we will show in the next subsection that $\Pi_{-1}$ vanishes for most tokamaks when momentum is not injected into the plasma due to a lowest order symmetry of the turbulence and the neoclassical transport.

\subsection{Up-down symmetry of the flux of toroidal angular momentum} \label{subsection:symmetry}

The lowest order neoclassical and gyrokinetic equations determine the functions $F_{i1}^\mathrm{nc} (\psi, \theta, v_{||}, \mu, t)$, $\underline{f}_{i1}^\mathrm{tb} (\psi, \theta, k_\psi, k_\alpha, v_{||}, \mu, t)$ and $\underline{\phi}_1^\mathrm{tb} (\psi, \theta, k_\psi, k_\alpha, t)$, where the velocity space is described by the parallel velocity $v_{||}$ and the magnetic moment $\mu = v_\bot^2/2B$. These functions depend on the local density and temperature gradients, the local rotation $\Omega_\zeta$, and its gradient $\partial \Omega_\zeta/\partial \psi$ at the location $\psi$. The functions $F_{i1}^\mathrm{nc} (\psi, \theta, v_{||}, \mu, t)$, $\underline{f}_{i1}^\mathrm{tb} (\psi, \theta, k_\psi, k_\alpha, v_{||}, \mu, t)$ and $\underline{\phi}_1^\mathrm{tb} (\psi, \theta, k_\psi, k_\alpha, t)$ only depend on the local parameters because the characteristic size of the turbulence and the width of the drift orbits that lead to neoclassical transport are small compared to the size of the machine. 

The rotation and the rotation gradient appear in the lowest order equations for $F_{i1}^\mathrm{nc} (\psi, \theta, v_{||}, \mu, t)$, $\underline{f}_{i1}^\mathrm{tb} (\psi, \theta, k_\psi, k_\alpha, v_{||}, \mu, t)$ and $\underline{\phi}_1^\mathrm{tb} (\psi, \theta, k_\psi, k_\alpha, t)$ only if the rotation is ordered to be sonic, $R\Omega_\zeta \sim v_{ti}$. This limit is the high flow ordering \cite{hinton85, artun94, sugama97, peeters09a}, and it has been adopted by many of the numerical studies of turbulent momentum transport performed so far \cite{waltz07, peeters07, peeters09b, casson09, casson10, barnes11a, highcock10}. If the rotation is sonic, most of the terms in \eq{eq:velocity} and \eq{eq:rotation} can be neglected to lowest order in $\rho_\ast$, leaving only $\bV_i = \Omega_\zeta R \zun$, with $\Omega_\zeta = - c (\partial \phi_0/\partial \psi)$. When the high flow ordering is employed, the lowest order equations for $F_{i1}^\mathrm{nc} (\psi, \theta, v_{||}, \mu, t)$, $\underline{f}_{i1}^\mathrm{tb} (\psi, \theta, k_\psi, k_\alpha, v_{||}, \mu, t)$ and $\underline{\phi}_1^\mathrm{tb} (\psi, \theta, k_\psi, k_\alpha, t)$ are invariant under the following transformation in an up-down symmetric flux surface: the sign of the rotation $\Omega_\zeta$ and the rotation shear $\partial \Omega_\zeta/\partial \psi$ is reversed, the sign of the variables $v_{||}$, $\mu$ and $k_\psi$ is reversed, and the sign of the functions $F_{i1}^\mathrm{nc}$, $\underline{f}_{i1}^\mathrm{tb}$ and $\underline{\phi}_1^\mathrm{tb}$ is reversed, i.e.
\begin{eqnarray} \label{eq:symmetry neoclassical}
F_{i1}^\mathrm{nc} (\psi, \theta, v_{||}, \mu, t; \Omega_\zeta, \partial \Omega_\zeta/\partial \psi) \rightarrow \nonumber\\ - F_{i1}^\mathrm{nc} (\psi, - \theta, - v_{||}, \mu, t; - \Omega_\zeta, - \partial \Omega_\zeta/\partial \psi),
\end{eqnarray} 
\begin{eqnarray} \label{eq:symmetry f turbulent}
\underline{f}_{i1}^\mathrm{tb} (\psi, \theta, k_\psi, k_\alpha, v_{||}, \mu, t; \Omega_\zeta, \partial \Omega_\zeta/\partial \psi) \rightarrow \nonumber\\ - \underline{f}_{i1}^\mathrm{tb} (\psi, - \theta, - k_\psi, k_\alpha, - v_{||}, \mu, t; - \Omega_\zeta, - \partial \Omega_\zeta/\partial \psi)
\end{eqnarray}
and
\begin{eqnarray} \label{eq:symmetry phi turbulent}
\underline{\phi}_1^\mathrm{tb} (\psi, \theta, k_\psi, k_\alpha, t; \Omega_\zeta, \partial \Omega_\zeta/\partial \psi) \rightarrow \nonumber\\ - \underline{\phi}_1^\mathrm{tb} (\psi, - \theta, - k_\psi, k_\alpha, t; - \Omega_\zeta, - \partial \Omega_\zeta/\partial \psi).
\end{eqnarray}
Symmetry \eq{eq:symmetry neoclassical} can be found in Appendix A of \cite{parra10a}, and symmetries \eq{eq:symmetry f turbulent} and \eq{eq:symmetry phi turbulent} were discussed in \cite{parra11c}. When these symmetries are applied to the lowest order piece of the momentum transport $\Pi_{-1}$ in \eq{eq:pi-1}, we find that in up-down symmetric flux surfaces, $\Pi_{-1}$ is odd in $\Omega_\zeta$ and $\partial \Omega_\zeta/\partial \psi$,
\begin{equation} \label{eq:pi odd}
\Pi_{-1} ( \Omega_\zeta, \partial \Omega_\zeta/\partial \psi ) = - \Pi_{-1} ( - \Omega_\zeta, - \partial \Omega_\zeta/\partial \psi ),
\end{equation}
leading to
\begin{equation} \label{eq:pi null}
\Pi_{-1} ( \Omega_\zeta = 0, \partial \Omega_\zeta/\partial \psi = 0 ) = 0.
\end{equation}
This symmetry is broken by up-down asymmetry in the flux surface \cite{camenen09a, camenen09b}, a fact that has been checked experimentally \cite{camenen10}. The effect of asymmetry is small in most existing tokamaks, particularly in the core, where the magnetic flux surfaces are very up-down symmetric.

\subsection{Consequences for intrinsic rotation} \label{subsection:consequences}

Equation \eq{eq:pi null} shows that there is no momentum flux in the absence of rotation, that is, without preexisting rotation, the turbulence or neoclassical effects will not redistribute the momentum. Another way of looking at this is that a solution for $\Pi_{-1} ( \Omega_\zeta, \partial \Omega_\zeta/\partial \psi) = 0$ with the boundary condition $\Omega_\zeta = 0$ at the edge is simply $\Omega_\zeta (\psi) = 0$ everywhere. There is no preferred direction and the plasma does not know in which direction it should rotate. This lack of preferred direction is only true for the high flow ordering. When we allow for a more general ordering below, we will be able to redistribute momentum even when there is no initial rotation.

If the boundary condition is not $\Omega_\zeta = 0$ at the edge, the solution to  $\Pi_{-1} ( \Omega_\zeta, \partial \Omega_\zeta/\partial \psi) = 0$ can be some non-trivial rotation. However, the profiles allowed by a flux of momentum odd in rotation and rotation shear are very limited, and will only depend on the condition at the edge. To see this, we Taylor expand around $\Omega_\zeta = 0$ and $\partial \Omega_\zeta/\partial \psi = 0$ to obtain
\begin{equation}
\Pi_{-1} \simeq - P \Omega_\zeta - \chi \frac{\partial \Omega_\zeta}{\partial \psi} = 0.
\end{equation}
Note that physically the high flow ordering only has diffusion and a pinch \cite{peeters07} in up-down symmetric configurations.  Solutions to this equation are always of the form
\begin{equation} \label{eq:solution high flow}
\Omega_\zeta (\psi) \propto \exp \left ( - \int \dd\psi\, \frac{P(\psi)}{\chi(\psi)} \right ).
\end{equation}
This rotation profile never changes sign from co-current to counter-current, making the high flow model unable to predict many observed profiles that change sign \cite{degrassie07, parra11e}. Another undesirable feature of solutions of the type \eq{eq:solution high flow} is that they depend completely on the boundary condition at the edge. This contradicts observations such as the rotation reversals \cite{bortolon06, rice11b} in which  a small change in the density leads to a jump from a completely co-current profile to a profile in which the rotation at the magnetic axis is counter-current and the rotation at the edge remains the same, that is, co-current.

To summarize, the high flow ordering, for which $R\Omega_\zeta \sim v_{ti}$, gives $\Pi_{-1} \neq 0$ and hence makes this piece of momentum transport the dominant contribution. However, it is not possible to explain the complex dependences observed in experiments with the results obtained with the high flow ordering. The reason is that the high flow ordering in up-down symmetric configurations only contains diffusion and a pinch, and therefore lacks any redistribution of momentum that is not induced by the presence of preexisting velocity.

The inability of the high flow ordering to cope with the intrinsic rotation problem indicates that we need to allow the velocity to be $R\Omega_\zeta \sim \rho_\ast v_{ti}$. With this low flow ordering, $\Omega_\zeta$ and $\partial \Omega_\zeta/\partial \psi$ do not enter in the equations for the lowest order pieces $F_{i1}^\mathrm{nc} (\psi, \theta, v_{||}, \mu, t)$, $\underline{f}_{i1}^\mathrm{tb} (\psi, \theta, k_\psi, k_\alpha, v_{||}, \mu, t)$ and $\underline{\phi}_1^\mathrm{tb} (\psi, \theta, k_\psi, k_\alpha, t)$, and $\Pi_{-1}$ is identically zero. As a result, $\Pi_0$ must be calculated. The formalism of \cite{sugama11} proves elegantly that $\Pi_0$ is unaffected by the symmetry that made $\Pi_{-1} = 0$, and a contribution to the momentum transport, $\Pi_\mathrm{int}$, is independent of rotation \cite{parra11d}, giving
\begin{equation}
\Pi_0 = - P \Omega_\zeta - \chi \frac{\partial \Omega_\zeta}{\partial \psi} + \Pi_\mathrm{int} \left ( \frac{\partial n_e}{\partial \psi}, \frac{\partial T_i}{\partial \psi}, \frac{\partial T_e}{\partial \psi}, \ldots \right ).
\end{equation}
Note that the solution to $\Pi_0 = 0$ clearly gives non-trivial solutions for $\Omega_\zeta (\psi)$. 

Thus, to correctly model intrinsic rotation, we need to obtain the contribution $\Pi_0 \sim \rho_\ast^3 p_i R |\nabla \psi|$, and this implies obtaining the second order pieces $F_{i2}$, $f_{i2}^\mathrm{tb}$ and $\phi_2^\mathrm{tb}$. Several authors have made some attempts at calculating this higher order contribution to momentum transport.  In \cite{parra10a, parra11d} a systematic expansion based on a subsidiary expansion in $B_p/B \ll 1$ is carried out. This subsidiary expansion assumes that the turbulent eddies and the size of the fluctuations does not scale strongly with $B/B_p$, and this simplifies the higher order equations enormously. Other authors have assumed that the main effect that contributes to $\Pi_0$ is the modification of the turbulence by the radial variation of the background gradients \cite{wang10, waltz11, camenen11}. Keeping only the radial variation of the background gradients means neglecting many other different effects that are as important as the one retained. Only by treating all the relevant terms on equal footing will it be possible to correctly reproduce the experimentally observed dependences. An extension of the work done in \cite{parra10a, parra11d} that relaxes the assumptions on the turbulence but still relies on the $B_p/B \ll 1$ expansion will be published shortly. This new improved model will include the effects already discovered in \cite{parra10a, parra11d}, and in addition will have the radial variation of the gradients and other effects that have not been considered before.

\section{Momentum transport in gyrokinetics} \label{section:fullf}

From the results in the previous section, it is clear that to correctly predict intrinsic rotation, the radial flux of toroidal angular momentum must be calculated up to order $\rho_\ast^3 p_i R |\nabla \psi|$. To do so, it is necessary to evaluate the next order correction to the fluctuations $f_{i2}^\mathrm{tb}$ and $\phi_2^\mathrm{tb}$. These corrections are small compared to what is usually calculated in gyrokinetic simulations by one factor of $\rho_\ast \ll 1$. 

A transport simulation that is based on fluid equations to evolve profiles can use the radial flux of toroidal angular momentum in \eq{eq:pi0} and then evaluate the turbulent fluctuations up to second order. To obtain the second order pieces, current gyrokinetic simulations have to be modified to include higher order effects such as those calculated in \cite{parra10a, parra11d}. These improvements are difficult to implement, but still tractable.

Transport codes that evolve the full distribution function and then obtain density, temperature and velocity profiles from moments of the distributions function require much more work to recover the correct intrinsic rotation profile. In \cite{parra10b} the order of magnitude estimate $\Pi \sim \rho_\ast^3 p_i R |\nabla \psi|$ was used to argue that these types of simulations need to obtain the gyrokinetic equation to order $\rho_\ast^4 f_i v_{ti}/L$, that is, to three orders of magnitude higher than it is usually calculated (only terms of order $\rho_\ast f_i v_{ti}/L$ are retained in most gyrokinetic full $f$ simulations). The reasons for this estimate are subtle (see \cite{parra10b}), but we can give an approximate idea here. The electrostatic potential is calculated using a gyrokinetic quasineutrality equation that naturally contains a polarization density,
\begin{equation} \label{eq:sketch quasineutrality}
\nabla \cdot \left ( \frac{Zcn_i}{B\Omega_i} \nabla_\bot \phi \right ) + \ldots = - Z \int \dd^3v\, f_{ig} + \int \dd^3v\, f_{eg}.
\end{equation}
In this equation we have kept a schematic polarization density on the left side, and the integrals over the distribution functions of the guiding centers of electrons, $f_{eg}$, and ions, $f_{ig}$, on the right side. The distribution functions of guiding centers are evolved using gyrokinetic equations. This equation is only meant to be schematic and it is not intended to be exhaustive. It will only serve for an order of magnitude estimate. In full $f$ simulations, the long wavelength electrostatic potential $\phi_0$ is to be obtained using equation \eq{eq:sketch quasineutrality}. From \eq{eq:velocity} and \eq{eq:rotation} we find that $\bV_i \sim (c/B) \nabla_\bot \phi_0$. Then, the time derivative of the polarization density due to the background potential $\phi_0$ is given by
\begin{equation}
\frac{\partial}{\partial t} \left [\nabla \cdot \left ( \frac{Zcn_i}{B\Omega_i} \nabla_\bot \phi_0 \right )  \right ] \sim \frac{1}{L} \left ( \frac{c n_i}{e B} m_i \frac{\partial \bV_i}{\partial t} \right ),
\end{equation}
and using the momentum transport equation \eq{eq:momentum transport} to make the order of magnitude estimate $n_i m_i (\partial \bV_i/\partial t) \sim R^{-1} \partial \Pi/\partial \psi \sim \rho_\ast^3 p_i/L$, we finally find
\begin{equation}
\frac{\partial}{\partial t} \left [\nabla \cdot \left ( \frac{Zcn_i}{B\Omega_i} \nabla_\bot \phi_0 \right )  \right ] \sim \frac{1}{L} \left ( \frac{c}{e B} \rho_\ast^3 \frac{p_i}{L} \right ) \sim \rho_\ast^4 n_i \frac{v_{ti}}{L}.
\end{equation}
Using this estimate in \eq{eq:sketch quasineutrality} leads to
\begin{equation}  \label{eq:size vorticity}
Z\int \dd^3v\, \frac{\partial f_{ig}}{\partial t} - \int \dd^3v\, \frac{\partial f_{eg}}{\partial t} \sim \rho_\ast^4 n_i \frac{v_{ti}}{L}.
\end{equation}
This condition implies that the gyrokinetic equations that give $\partial f_{ig}/\partial t$ and $\partial f_{eg}/\partial t$ must be obtained up to order $\rho_\ast^4 f_i v_{ti}/L$. This requirement makes the use of codes that are not based on fluid equations to model intrinsic rotation an almost impossible task.

The method that is most widely used to derive the gyrokinetic equation for full $f$ simulations is the phase-space Lagrangian approach \cite{brizard07}. For this reason, we now proceed to discuss the implications of what we have derived so far for Lagrangian methods. In the Lagrangian approach to the derivation of gyrokinetics, a coordinate transformation is performed in the phase-space Lagrangian of a particle in an electromagnetic field. The idea is obtaining a phase-space Lagrangian independent of gyrophase. This cannot be done exactly, but the gyrophase dependence can be eliminated to any order of interest (although the work required increases rapidly with the orders). Once the phase-space Lagrangian is calculated, a variational approach \cite{sugama00, brizard00b} is used to obtain the equations of motion for the particles and Maxwell's equations for the electromagnetic fields. This variational method makes it necessary to have a Hamiltonian correct to order $\rho_\ast^3 T_e$ to obtain a quasineutrality equation sufficiently accurate to obtain $f_{i2}^\mathrm{tb}$ and $\phi_2^\mathrm{tb}$. The need for a Hamiltonian of this order to calculate the second order pieces $f_{i2}^\mathrm{tb}$ and $\phi_2^\mathrm{tb}$ has been argued for a slab in \cite{parra10b} and for a tokamak in \cite{parra11a}. 

The question of the order to which the Hamiltonian is needed is not completely resolved by this argument. If the gyrokinetic equation is used only to calculate the fluctuations, which are then employed to evaluate the radial turbulent flux of momentum \eq{eq:pi0}, we would not need to consider anything further. However, if the variational formulation is used to evolve the full distribution function and then the velocity is evaluated by taking a moment, we need to check if the gyrokinetic equation obtained with the Hamiltonian up to order $\rho_\ast^3 T_e$ is sufficient to recover the correct transport of momentum. The answer is not obvious since the fourth order Hamiltonian gives corrections to the gyrokinetic equation that are of order $\rho_\ast^4 f_i v_{ti}/L$, and these may matter for momentum transport, as we saw in \eq{eq:size vorticity}. In \cite{parra10b} it was proven that the Hamiltonian up to order $\rho_\ast^3 T_e$ was sufficient in a slab. It is possible to show that it is also sufficient for a tokamak. By using the variational Lagrangian formulations, it is possible to obtain conservation equations that look similar to the conservation of toroidal angular momentum \cite{scott10, brizard11}. The actual physical conservation of toroidal angular momentum and these gyrokinetic conservation equations are not equal to each other, but they tend to the same equation at long wavelengths. This property is what makes it possible to prove that the Hamiltonian to order $\rho_\ast^3 T_e$ is sufficient to obtain the correct rotation. The calculation that has achieved the highest order Hamiltonian is \cite{parra11a}, and the Hamiltonian in this reference is only accurate to order $\rho_\ast^2 T_e$. Thus, it is still necessary to calculate a higher order correction to the Hamiltonian. In addition, the conservation properties derived so far did not consider collisions. Collisions are important for the radial flux of momentum transport even for very low collisionality because there is a piece of $F_{i2}$ that is turbulent in nature and depends on the collision operator \cite{parra10a, parra11d}. This piece will give a contribution to $\Pi_0$ in \eq{eq:pi0} as large as any of the other turbulent contributions. The treatment of the collision operator in full $f$ simulations remains a challenge.

\section{Discussion} \label{section:discussion}

It is clear that to reproduce the variety of behaviors observed in intrinsic rotation experiments with gyrokinetic simulations, it is necessary to retain many effects that are small in $\rho_\ast$ compared to the lowest order gyrokinetic formulation. These effects that are in principle small in $\rho_\ast$ become the dominant contribution because of a cancellation of the lowest order momentum flux due to a symmetry in the system. Many new contributions that have been previously neglected now become important. To attain predictive capability, we need to let the different contributions compete to determine the rotation direction. The problem is very complex, but it can be simplified using a subsidiary expansion in $B_p/B \ll 1$ \cite{parra10a, parra11d}. This expansion has allowed us to prove that the intrinsic rotation will depend on the density and temperature gradients \cite{parra11d}, and it has given us a characteristic size of the intrinsic rotation generated by the turbulence, $\bV_i \sim (B/B_p) \rho_\ast v_{ti}$. This order of magnitude estimate gives a clear prediction for the scaling of the intrinsic rotation generated in the core of tokamaks with temperature and plasma current that has been confirmed by comparing data from several machines \cite{parra11e}.

The implementation of the new low flow formulation of gyrokinetics necessary to describe intrinsic rotation is challenging, but it is tractable for existing $\delta f$ simulations. The same cannot be said for gyrokinetic simulations that evolve the full distribution function. These simulations need a gyrokinetic equation derived to high order in $\rho_\ast$, and they cannot benefit from many of the simplifications that make $\delta f$ simulations computationally faster and more manageable. The required high order gyrokinetic equation can be obtained in the collisionless limit if the gyrokinetic Hamiltonian is obtained to order $\rho_\ast^3 T_e$. The addition of collisions to these full $f$ models remains unresolved. It is very likely that the full $f$ gyrokinetic formulation can be simplified with a subsidiary expansion in $B_p/B \ll 1$. However, even then the numerical accuracy to which the gyrokinetic equation needs to be solved is extremely high, and scales like $\rho_\ast^4$, as indicated by the requirement in \eq{eq:size vorticity}. Numerical methods that ensure this accuracy need to be considered. Wherever the scale length $L$ of the phenomena of interest is small (e.g. transport barriers), the accuracy constraint $\rho_\ast^4 = (\rho_i/L)^4$ may not seem as daunting, but it remains a condition to be dealt with, in particular regarding the equations, which need to be derived to this order.

\acknowledgments The authors thank G. Hammett and J. Krommes for many helpful discussions. This research was supported in part by the U.S. Department of Energy Grant No. DE-FG02-91ER-54109 at the Plasma Science and Fusion Center of the Massachusetts Institute of Technology and by the grant ENE2009-07247, Ministerio de Ciencia e Innovaci\'{o}n (Spain).

\end{document}